\documentclass[fleqn,usenatbib]{mnras}

\usepackage{newtxtext,newtxmath}
\usepackage{algorithmic}
\usepackage[T1]{fontenc}
\usepackage{ae,aecompl}

\usepackage{graphicx}	% Including figure files
\usepackage{amsmath}	% Advanced maths commands
\usepackage{amssymb}	% Extra maths symbols

\title[Calibration of a Low-Frequency SKA-station]{On-sky calibration of a SKA1-low station in the presence of mutual coupling}

\author[J. Borg et al.]{J. Borg,$^{1}$\thanks{E-mail: josef.borg@um.edu.mt}
A. Magro,$^{1,2}$
K. Zarb Adami,$^{1,2,3, 4}$
E. de lera Acedo,$^{5}$
\newauthor
A. Sutinjo,$^{6}$
and D. Ung$^{6}$
\\
$^{1}$Institute of Space Sciences and Astronomy, University of Malta, Msida MSD 2080, Malta\\
$^{2}$Department of Physics, University of Malta, Msida, MSD 2080, Malta\\
$^{3}$Department of Physics, University of Oxford, Denys Wilkinson Building, Keble Road, Oxford OX1 3RH. UK \\
$^{4}$Osservatorio Astrofisico di Catania, Via S. Sofia 78, 95123, Catania, Italy \\
$^{5}$Cavendish Astrophysics, University of Cambridge, JJ Thompson Ave, CB3 0HE, Cambridge, UK \\
$^{6}$International Centre for Radio Astronomy Research (ICRAR), Curtin University, Bentley, Western Australia
}

\date{Accepted XXX. Received YYY; in original form ZZZ}

\pubyear{2020}

\begin{document}
\label{firstpage}
\pagerange{\pageref{firstpage}--\pageref{lastpage}}
\maketitle

\begin{abstract}
This paper deals with the calibration of the analogue chains of a Square Kilometre Array (SKA) phased aperture array station, using embedded element patterns (one per antenna in the array, thus accounting for the full effects of mutual coupling) or average element patterns to generate model visibilities. The array is composed of 256 log-periodic dipole array antennas. A simulator capable of generating such per-baseline model visibility correlation matrices was implemented, which allowed for a direct comparison of calibration results using StEFCal (Statistically Efficient and Fast Calibration) with both pattern types. Calibrating the array with StEFCal using simulator-generated model visibilities was successful and thus constitutes a possible routine for calibration of an SKA phase aperture array station. In addition, results indicate that there was no significant advantage in calibrating with embedded element patterns, with StEFCal successfully retrieving similar per-element coefficients with model visibilities generated with either pattern type. This can be of significant importance for mitigating computational costs for calibration, particularly for the consideration of real-time calibration strategies. Data from the AAVS-1 (Aperture Array Verification System 1) prototype station in Western Australia was used for demonstration purposes. 
\end{abstract}

\begin{keywords}
instrumentation: interferometers -- techniques: interferometric -- methods: data analysis
\end{keywords}

%%%%%%%%%%%%%%%%%%%%%%%%%%%%%%%%%%%%%%%%%%%%%%%%%%

%%%%%%%%%%%%%%%%% BODY OF PAPER %%%%%%%%%%%%%%%%%%

\section{Introduction}

SKA1-low will be one of two telescopes to be deployed during Phase 1 of the Square Kilometre Array (SKA). The SKA project will be jointly hosted across two continents and will have a maximum total collecting area of 1 million square metres (\cite{schilizzi2008square}). SKA1-Mid will consist of an interferometer with 197 dishes (15 m in diameter) hosted in the Karoo radio reserve in South Africa, while SKA1-low will be hosted in Western Australia (\cite{dewdney2009square}). The latter will comprise of 512 stations (phased aperture arrays) of 256 log-periodic dipole array antennas each for a total of \ensuremath{2^{17}} antennas, covering the low frequency radio regime between 50 and 350 MHz (\cite{farnes2018science}). By the time of its completion the SKA will be the largest and most powerful radio telescope at m- and cm- wavelengths, with an expected sensitivity $\sim$50 times that of any other radio telescope (\cite{johnston2007science}). 

The site selected for SKA1-low is a radio-quiet zone in the Australian outback, spanning several thousand km\ensuremath{^{2}} of extremely sparsely populated land, making it ideal for faint radio signal observations due to very low levels of radio frequency interference (RFI) (\cite{offringa2015low}). With all antennas being stationary, SKA1-Low will truly be an all-electronic telescope enabled by advanced signal processing, with the telescope's sheer scale presenting a significant challenge. 

The main scientific targets for SKA1-low include high red-shift imaging of HI (\cite{datta2016probing}), pulsar timing (\cite{smits2009pulsar}) and detection of the Epoch of Re-ionization (EoR) signal (\cite{faulkner2015ska}, \cite{mellema2013reionization}). Due to the large number of sparsely located antennas, SKA1-low will be capable of highly flexible aperture control which can be easily configured using software and which shall allow very high dynamic range observations.

As part of the verification process, a single SKA1-low station of 256 antennas has been deployed in Murchison Radio Observatory (MRO) (\cite{tingay2013murchison}) in the Western Australian desert, named Aperture Array Verification System 1 (AAVS-1) (\citeauthor{aavs1benthem} (in prep.)). AAVS-1 follows the previous SKA verification system, AAVS-0.5, which was composed of 16 log-periodic dipole antennas (\cite{sutinjo2015characterization}). Although AAVS-1 could be used as a scientific instrument in its own right, the main purpose of the array is that of developing a prototype routine for SKA1-low station design validation. The development of a calibration strategy is an important aspect of this validation process, with every SKA1-low station required to carry out its own calibration routine accordingly. As such, AAVS-1 provides an ideal platform for assessing such strategies (\cite{magro2017monitoring}). 

This study proposes and verifies one such calibration strategy for AAVS-1 while also assessing the requirement for consideration of embedded pattern responses as a secondary objective. In itself, the ability to consider embedded element patterns is desired since it allows for generation of model visibilities accounting for the full effects of mutual coupling within the array. For this reason, the possibility to consider such effects was included in this calibration strategy accordingly. It can be noted, however, that with a total of \ensuremath{2^{17}} antennas grouped in 512 distinct stations, each with its own unique arrangement of antennas, mutual coupling will also vary from station to station. This effectively means that an embedded pattern could potentially need to be computed and stored for every antenna making up the final SKA1-Low telescope, per frequency channel, per polarization, in order to completely consider mutual coupling effects. This could amount to a requirement of approximately 50Tb of storage for the embedded patterns to be stored in spherical harmonics form, considering 512 stations of 256 dual-polarization antennas each for 400 channels across the 350MHz observation window for SKA-1 Low. This would also incur a computational run-time cost for spherical wave expansion to compute full embedded patterns in real-time. Conversely, if previously computed patterns are stored and loaded as necessary, this would still incur random access memory requirements of approximately 215GB per station, equating to a total of around 131Tb for all stations.

This is therefore of significant importance to the computational cost considerations for SKA1-Low station calibration since it constitutes a direct impact on routine efficiency, in particular, during simulation of model visibilities. Indeed, consideration of embedded element patterns would negatively effect calibration routine run-time, whether such patterns are generated using the spherical wave expansion or previously computed and loaded for model simulations. For the possibility of real-time calibration on a per-frequency channel basis for such arrays, optimization of calibration routine efficiency would be crucial.

The calibration strategy developed in this study required the implementation of an observation simulator, capable of generating model visibilities for a phased array observation using an appropriate sky model and the required element patterns. Therefore, the simulator here introduced is able to fully consider mutual coupling effects within the AAVS-1 station, provided appropriate pattern information. Calibration with StEFCal (\textbf{St}atistically \textbf{E}fficient and \textbf{F}ast \textbf{Cal}ibration) (\cite{salvini2014fast}) was implemented, using the simulator-generated per-baseline visibilities to retrieve per-element calibration coefficients for channelized observation data retrieved from AAVS-1. In this manner, this study aimed to ascertain the ability to calibrate AAVS-1, a SKA1-low precursor array, using different calibration observation coordinates. As a secondary objective, calibration using visibility correlation matrices generated using embedded element pattern responses was assessed against the same calibration routine using visibilities with only an average element pattern response considered. In this manner, the study aimed to also ascertain whether embedded element patterns are a necessary consideration for calibrating an AAVS-1 station.

\section{Simulations and Observations}

The simulator and calibration strategy implemented will be hereunder described. Section \ref{MVS} separately deals with the simulation and perusal of element patterns (subsection \ref{EPS}), the selection of a local sky model from a global sky model, relevant for a particular observation (subsection \ref{LSM}) and the computation of per-baseline visibilities to populate a model correlation matrix {subsection \ref{MVC}}. Section \ref{data_acq} identifies data acquisition details, using AAVS-1, which was hence used for calibration testing. Subsequently, section \ref{CAL} identifies the calibration routine employed and presents calibration results obtained, inclusive of a comparison of results obtained with model visibilities generated using embedded or average element patterns. 

\subsection{Model Visibilities Simulator}\label{MVS}
\label{sec:maths} % used for referring to this section from elsewhere

The prototype simulator implemented for this study was developed in Python. It allows for several simulation parameters to be defined, making a number of different simulation scenarios possible. User-defined parameters include the telescope model itself, observation-specific parameters such as observation start time, end time and time step interval, observation frequency channel, sky model to be used and frequency-specific embedded element pattern data. In place of the latter, spherical harmonic data from which the simulator can generate embedded pattern responses at a specified frequency (\cite{de2011compact}) accordingly can be provided instead. 

% figure
\begin{figure}
	\includegraphics[width=\columnwidth]{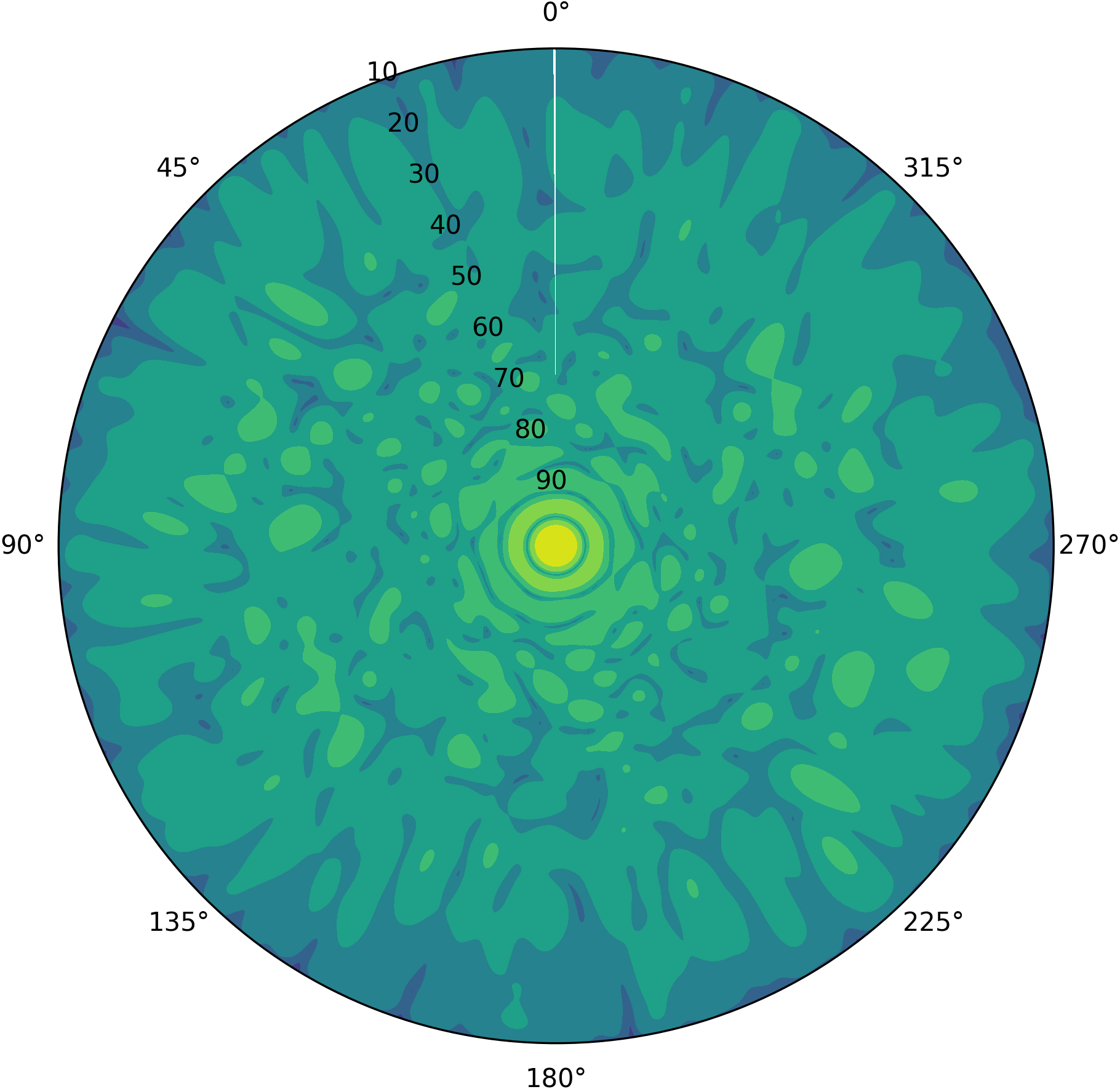}
	\includegraphics[width=\columnwidth]{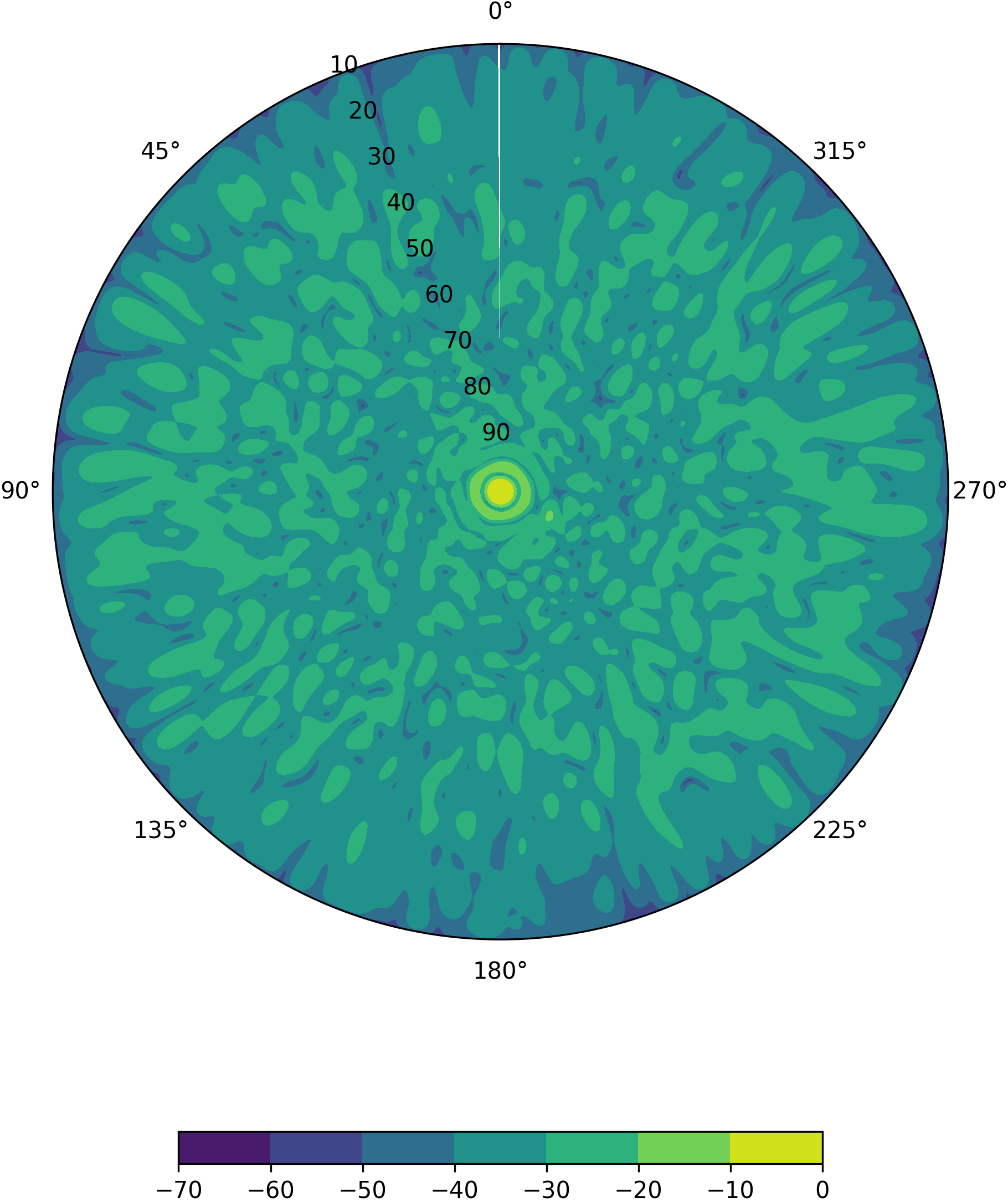}
    \caption{Normalized station beam pattern power (in dB) for an AAVS-1 station of 256 elements, using embedded element patterns retrieved from spherical harmonics using the spherical wave expansion at 110MHz (above) and 160MHz (below).}
    \label{fig:station_pattern}
\end{figure}

Additionally, the simulator can be configured to generate an average element pattern from the embedded element patterns, either for testing purposes or for requested calibration routine runs. It is possible to generate observation-specific sky models for an observation, such that the user can generate the required models to calibrate with from the array's location at the observation time. It is however also possible to use sky models previously generated with the simulator, thus resulting in less computation time required for generation of user-requested per-baseline visibilities.

\subsubsection{Element Pattern Simulations}\label{EPS}

The simulator can make use of either previously generated embedded pattern responses or generate its own embedded element responses from spherical harmonics provided by the user, as aforementioned. The latter allows for multi-frequency simulations requiring a significantly smaller storage footprint per frequency channel, without a loss in accuracy in obtaining arbitrary far-field points for a beam model. This, however, incurs a computation time penalty, which could be a problem for real-time calibration routines. Figure \ref{fig:station_pattern} shows the AAVS-1 station pattern calculated as demonstrated in equation \ref{eq:LC_1} at 110MHz and 160MHz, combining embedded element patterns retrieved from spherical harmonics using the spherical wave expansion, as described in \cite{sokolowski2017calibration} and \cite{de2011compact}. 

The element pattern simulations in this paper have been performed using the commercial code FEKO (\cite{feko00}) and validated against the in-house code HARP (\cite{bui2018fast}). Both of these employ methods of moments electromagnetic solvers, but they use significantly different approaches for the solution of the array mutual coupling.

\subsubsection{Local Sky Model Selection}\label{LSM}

The local sky model refers to the hemisphere from a global sky model which is present above the horizon at an observation time \ensuremath{\tau_{0}} from a particular location on the Earth's surface. With a continuously changing local sky model, any simulated observation starting at \ensuremath{\tau_{0}} and ending at \ensuremath{\tau_{1}} with a time step interval of \textit{t} seconds will require \textit{n} different local sky models to be computed, with one model sky for every separate interval required accordingly. 

For calibration purposes, the time-step interval can be user-defined and would be expected to vary between observations, depending on the instrument, observation details and the observer's scientific goal. The time step interval would be selected according to the timescale over which coherence for a particular observation is expected to be lost, referred to as the coherence time (\cite{rogers1981coherence}). This would change significantly with varying array baseline lengths and observation frequencies, with longer baselines and higher frequencies reducing coherence time to possibly sub-minute timescales for certain scientific goals. The requirement for a fast, real-time calibration strategy could thus be required in such cases, with successive local sky model selection on short time scales being necessary for frequent calibration runs. 

% figure
\begin{figure}
    \centering
	\includegraphics[width=.8\columnwidth]{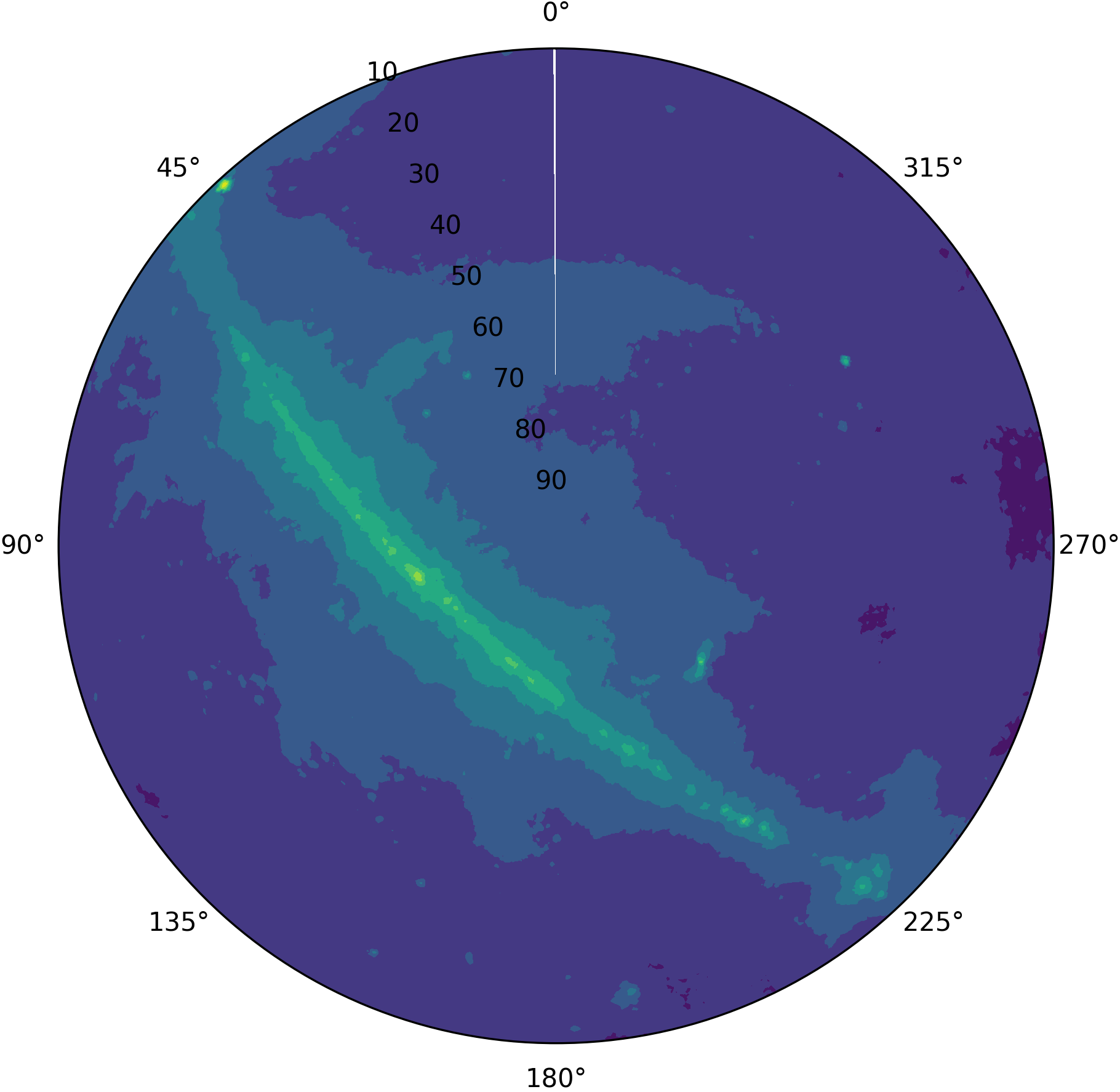}
	\includegraphics[width=.8\columnwidth]{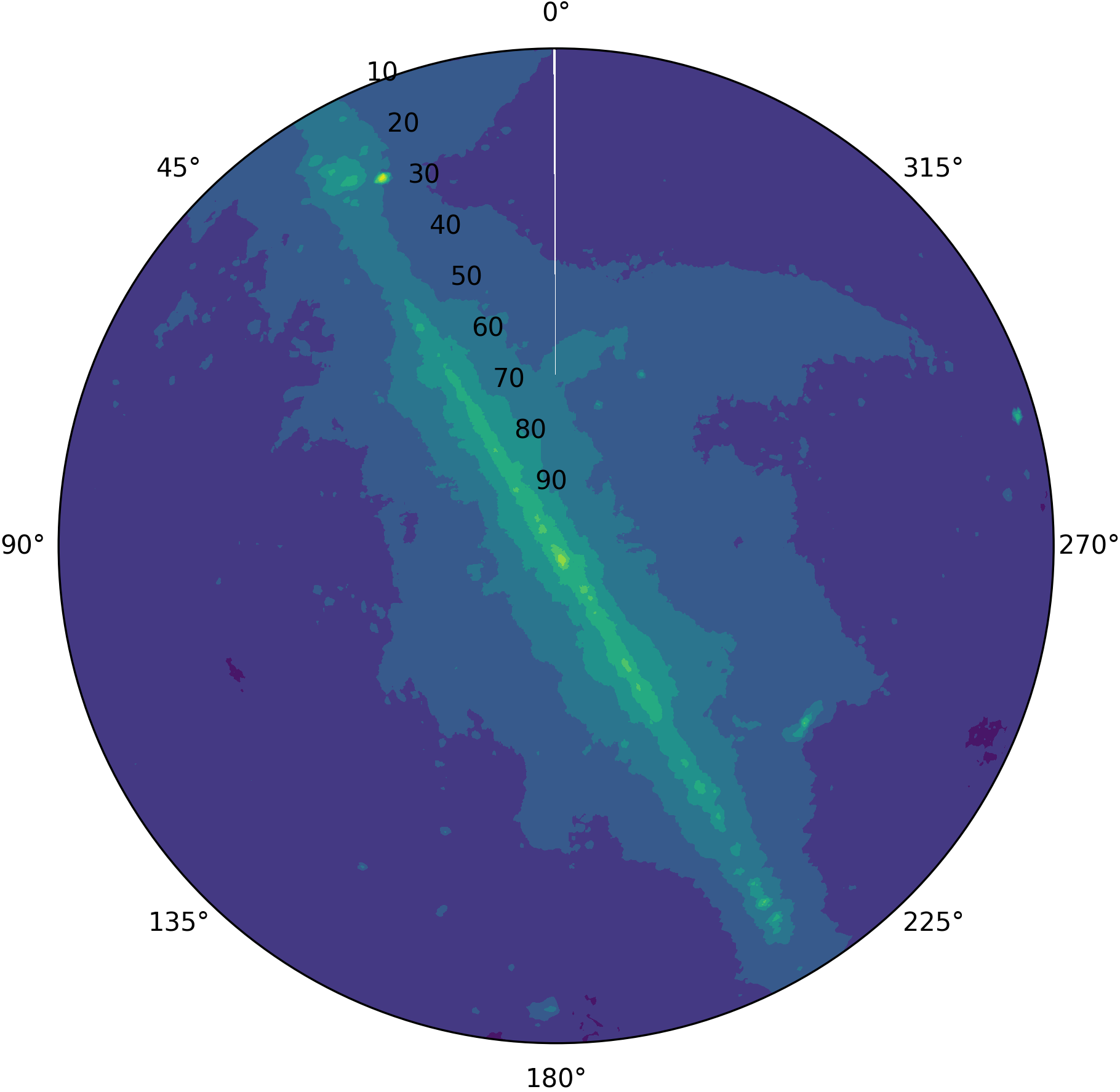}
	\includegraphics[width=.8\columnwidth]{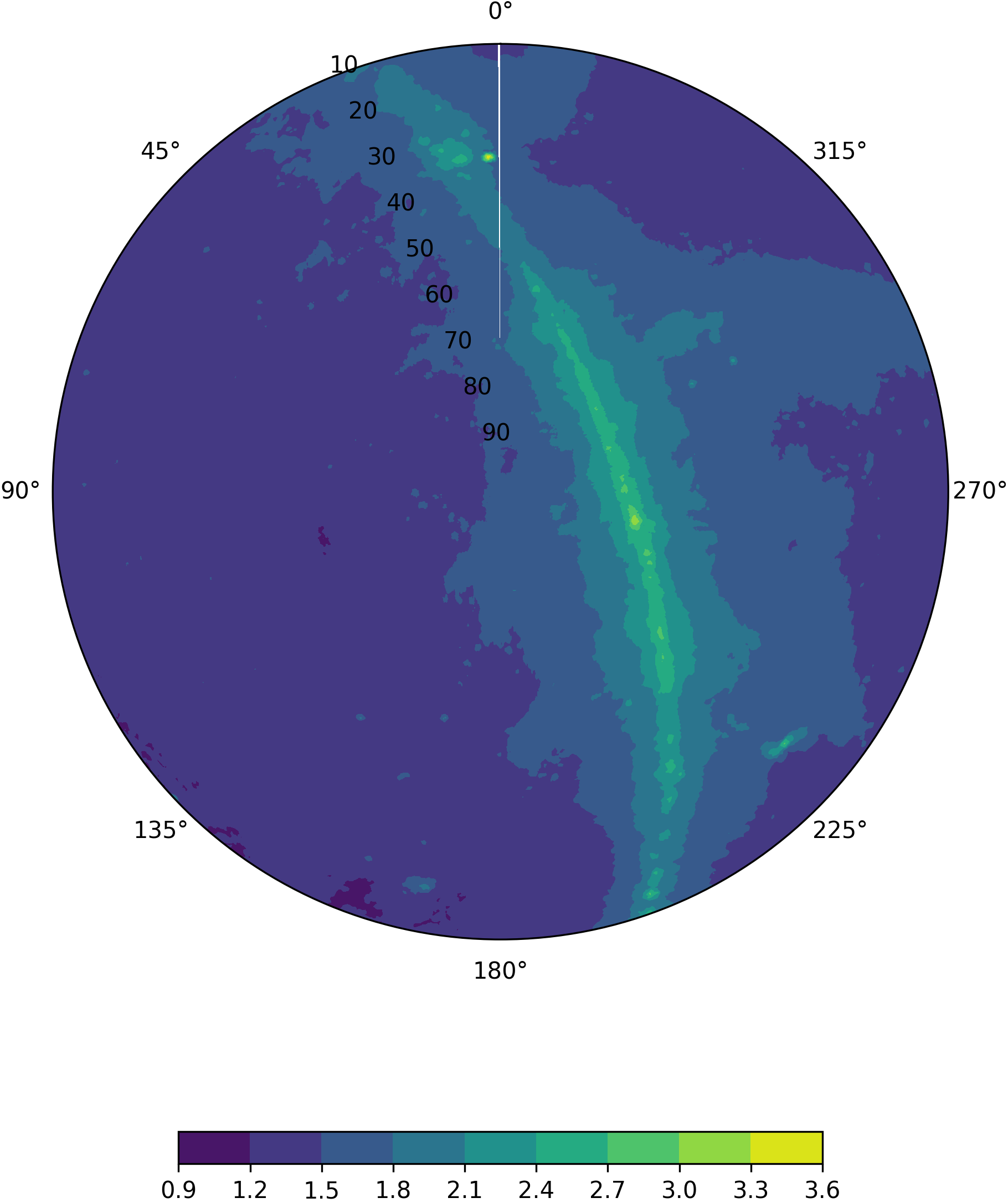}
    \caption{Reproduced snapshot projections of the local sky model from the AAVS-1 station location, showing logarithmic representations of brightness temperature. Snapshots for 2 hours prior to meridian transit (above), at meridian transit (middle) and 2 hours after meridian transit of Sagittarius A* (below) are shown.}
    \label{fig:sky_model}
\end{figure}

The global sky model used for the simulations carried out in this study is the 408MHz HASLAM All-Sky Map (\cite{haslam1982408}), scaled accordingly for simulations at different observation frequencies. Figure \ref{fig:sky_model} represents three local sky model projections at different times; two hours prior to, two hours after and at the meridian transit time for Sagittarius A*. The equatorial right ascension \ensuremath{\alpha} and declination \ensuremath{\sigma} coordinates of a point on the sky present above the telescope location with longitude \ensuremath{\omega} and latitude \ensuremath{\psi} at observation Julian date \ensuremath{\tau_{JD}} were calculated for every horizontal coordinate point with altitude \ensuremath{\theta} and azimuth \ensuremath{\phi}.

The declination \ensuremath{\sigma} can be calculated as shown in equation \ref{eq:SM_1}.

\begin{equation}
    \sigma = \arcsin(\sin\theta \sin\psi + \cos\theta \cos\psi cos\phi)
    \label{eq:SM_1}
\end{equation}

On the other hand, right ascension \ensuremath{\alpha} is defined as given in equation \ref{eq:SM_2}, where \ensuremath{\Theta_{\text{LST}}} is the local sidereal time, calculated from Greenwich sidereal time (GST) and the longitude \ensuremath{\omega}.

\begin{equation}
    \alpha = \Theta_{\text{LST}} - \arctan\Bigg(\frac{-\sin\phi \cos\theta / \cos\sigma}{\sin\theta - \sin\sigma \sin\psi / \cos\sigma \cos\psi}\Bigg) 
    \label{eq:SM_2}
\end{equation}

The simulator's local sky model selection module consequently selects the global sky model right ascension and declination sky brightness values located closest to the  equatorial coordinates above the horizon from the global HASLAM map sky model. The local sky model from the AAVS-1 location at an intended observation time was hence retrieved for calibration purposes in this study.

The local sky model selection would be expected to be a computationally expensive task to undergo for every different observation carried out. Hence, sky models can be generated once at a particular time step interval and saved for further perusal in future simulations with the same interval. Such sky models can be generated at a small time scale interval, equivalent to the minimum coherence time envisaged as a sufficient requirements for any planned observational scientific targets. In this manner, sky model selection and model visibilities generation would not be a requisite step for any real-time calibration strategies. Users could then peruse required model visibilities from a previously compiled model library, matching their observation parameters accordingly at the time interval necessary for their respective observation goals. This would mean that observers could progress directly to the chosen calibration routine, used to calibrate in real time with a frequency equivalent to their chosen time interval.

\subsubsection{Model Visibilities Computation}\label{MVC}

With the embedded element patterns and the local sky model computed at the same resolution and projected in the same coordinate system, pixels between the two were matched. The resolution selected for simulation purposes was that of the local sky model selected from the HASLAM map, since it is possible to generate the embedded element patterns at any required resolution from spherical harmonics. It was therefore possible to compute model baseline visibilities using equation \ref{eq:VIS_1} (\cite{ung2019modelling}, \cite{ung2020noise}).

\begin{equation}
    \big \langle v_{i} v_{j}^{*} \big \rangle = Z_{f} \frac{k}{\lambda^{2}}\big|Z_{\text{LNA}}\big|^{2} \Bigg{(}\frac{4\pi}{\omega \mu_{0}} \Bigg{)}^{2} c
    \label{eq:VIS_1}
\end{equation}

The term \textit{c} is the integral for the cross-correlations for antennas \textit{i} and \textit{j} across all sky directions, as shown in equation \ref{eq:VIS_2}, where \textit{T} is the sky temperature brightness at horizontal coordinates \ensuremath{\theta} and \ensuremath{\phi} and \ensuremath{\bar{e}^{i}} is the element pattern response for antenna \textit{i} in the same horizontal coordinate directions.

\begin{equation}
    c = \int\limits_{0}^{2\pi} \int\limits_{0}^{\pi} T(\theta, \phi) \Big{[}\bar{e}^{i}(\theta, \phi) \cdot \bar{e}^{j}(\theta, \phi)^{*}\Big{]} \sin{\theta} d\theta d\phi
    \label{eq:VIS_2}
\end{equation}

\begin{figure}
	\includegraphics[width=\columnwidth]{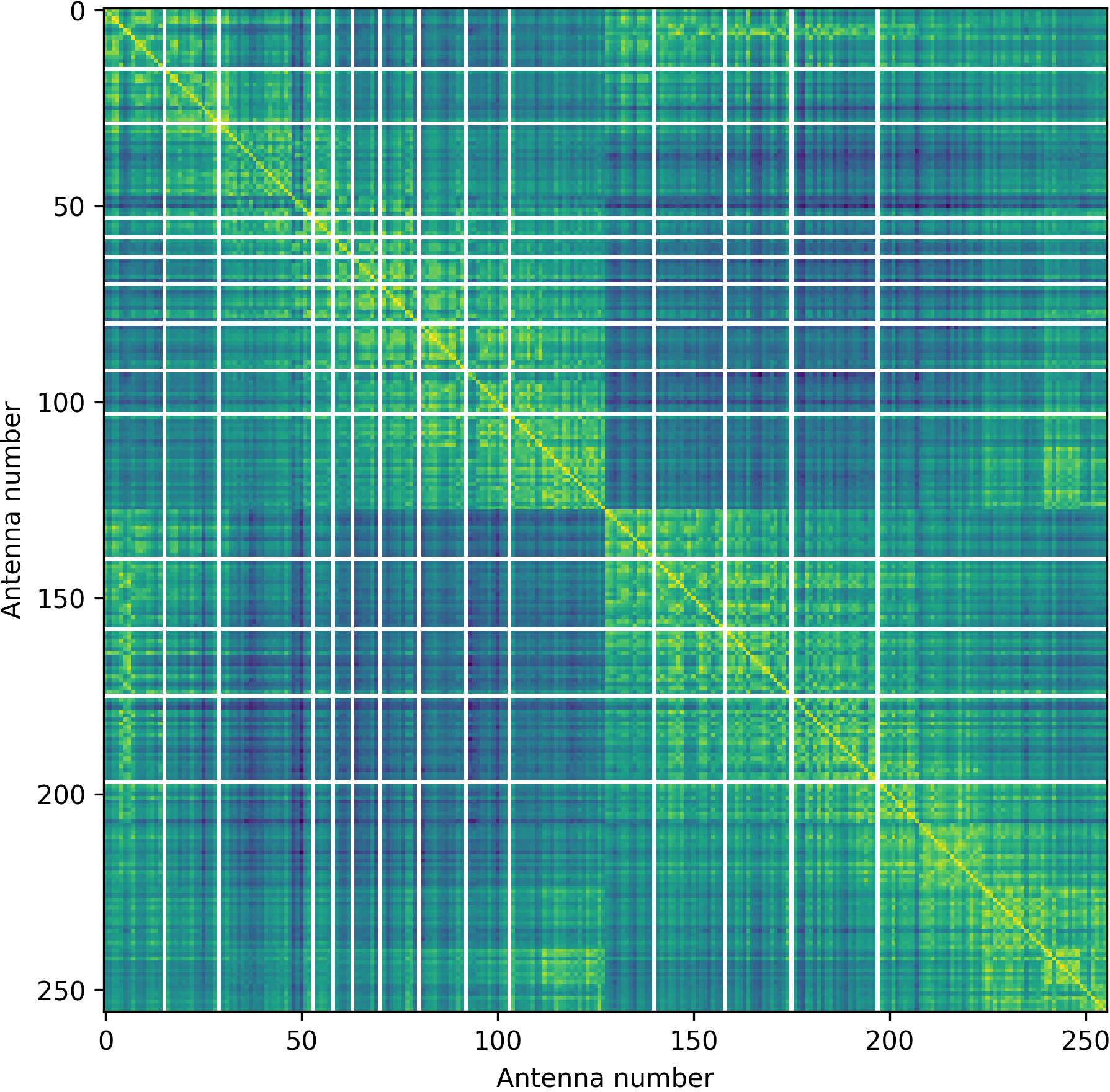}
    \caption{Model visibilities correlation matrix showing baseline power, with antenna numbers (0-255) subtending all respective baselines indicated. A number of user-defined antennas were switched off for this simulation and hence any baselines involving these antennas have zero power.}
    \label{fig:vis_corr}
\end{figure}

The simulator computes the full correlation matrix of model baseline visibilities, inclusive of auto-correlations. Figure \ref{fig:vis_corr} is a representation of a full correlation matrix of model visibilities, with some user-defined antennas switched off. This is a necessary consideration for any real-world calibration strategy, since inactive or damaged antennas in an array would otherwise be incorrectly calibrated, skewing complex gain coefficient computation for other array antennas.   

% figure

For calibration testing purposes, it is also possible to generate mock uncalibrated visibilities with added phase errors, also user-defined. Any phase error defined by the user is a range limit, and every element will be assigned a random Gaussian phase error within these set limits. These per-element errors are recorded to allow assessment of calibration testing results accordingly. Equation \ref{eq:VIS_3} shows the inclusion of complex noise terms in the computation of baseline visibilities, where \ensuremath{\big \langle Ve_{i} Ve_{j}^{*} \big \rangle} is the mock real cross-correlation for the baseline between antennas \textit{i} and \textit{j} with per-element phase errors \ensuremath{\phi_{i}} and \ensuremath{\phi_{j}} respectively. 
\begin{equation}
    \big \langle Ve_{i} Ve_{j}^{*} \big \rangle = \big \langle V_{i} V_{j}^{*} \big \rangle \cdot (\phi_{i}\phi_{j}^{*})
    \label{eq:VIS_3}
\end{equation}

% figure
\begin{figure*}
	\includegraphics[width=\textwidth]{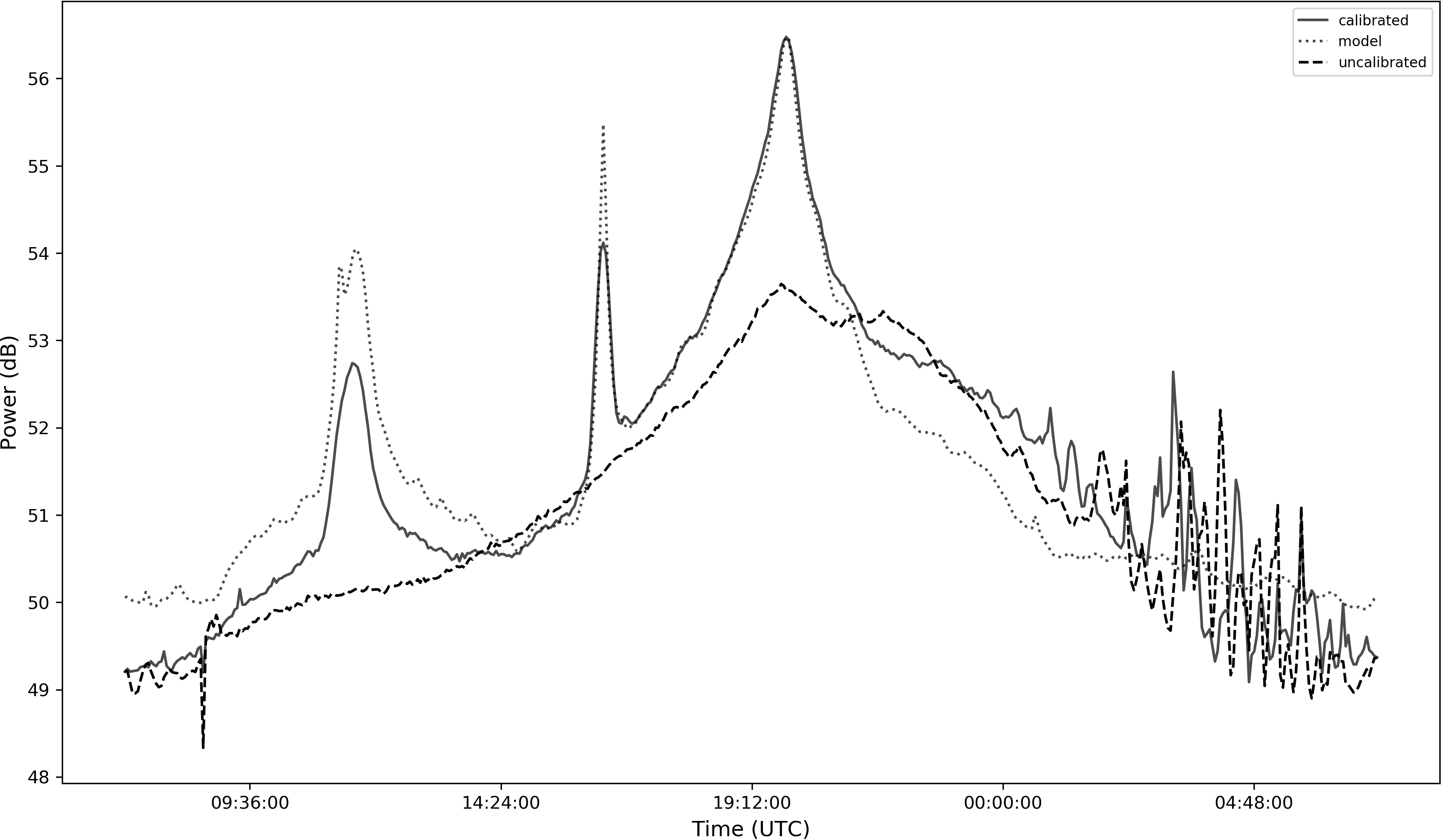}
    \caption{A 24 hour observation calibrated using model visibilities generated with an average element pattern, steered to the meridian transit coordinates for Centaurus A. Calibration was carried out separately at 900 second intervals, in order to calibrate on Sagittarius A* at different elevations, but only one of these calibrated data sets is shown in this figure (calibration with Sagittarius A* at highest elevation, during meridian transit). The three distinct peaks retrieved correspond to the expected passage times of the Vela-X pulsar, the Centaurus A galaxy and the Milky Way galactic plane (region in Scorpius) respectively. The 24 hour model visibilities used for calibration are included for comparison, as are the original uncalibrated visibilities for the 24 hour data set retrieved from AAVS-1.}
    \label{fig:cen_emb}
\end{figure*}

% figure
\begin{figure}
	\includegraphics[width=\columnwidth]{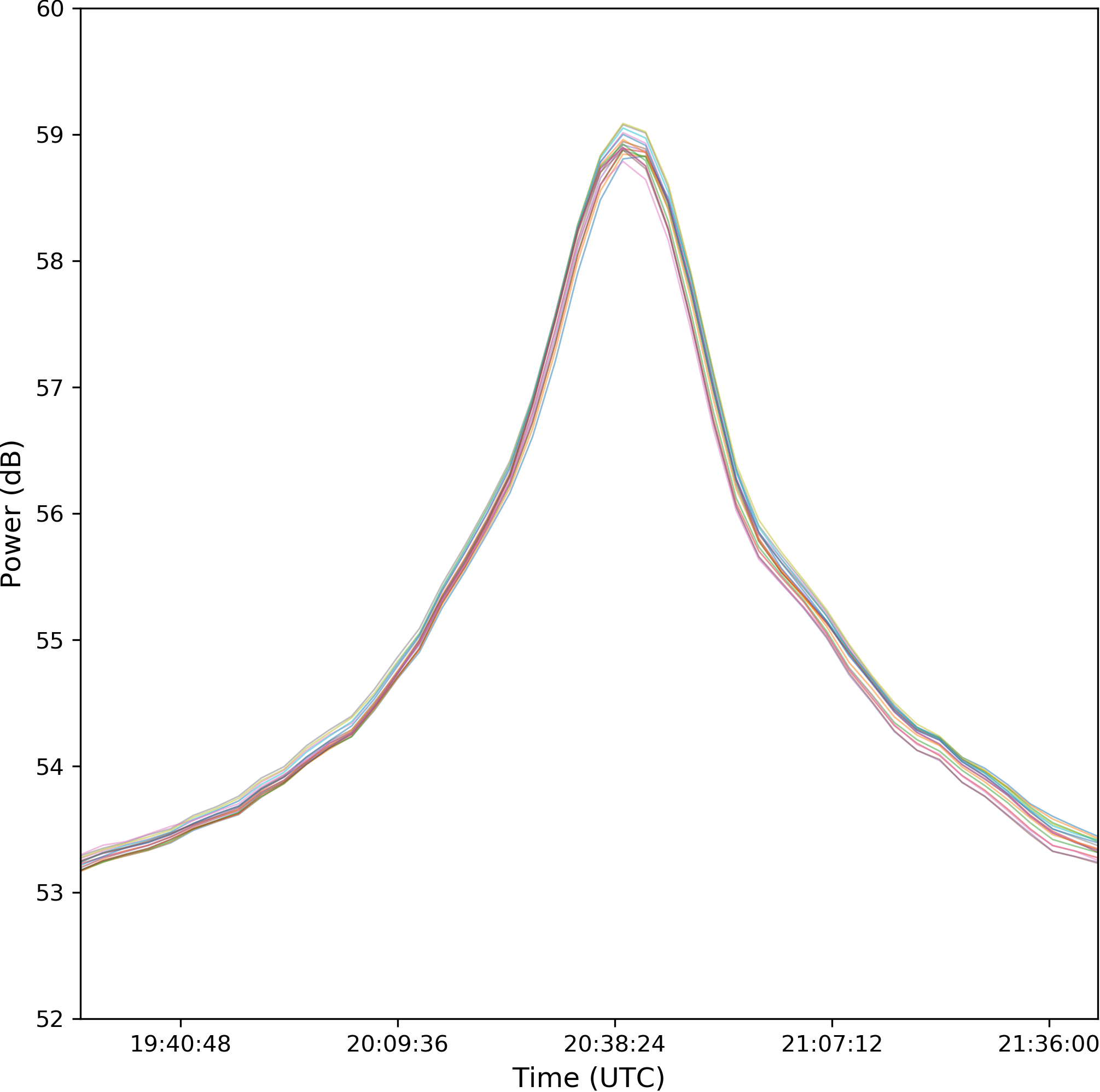}
	\includegraphics[width=\columnwidth]{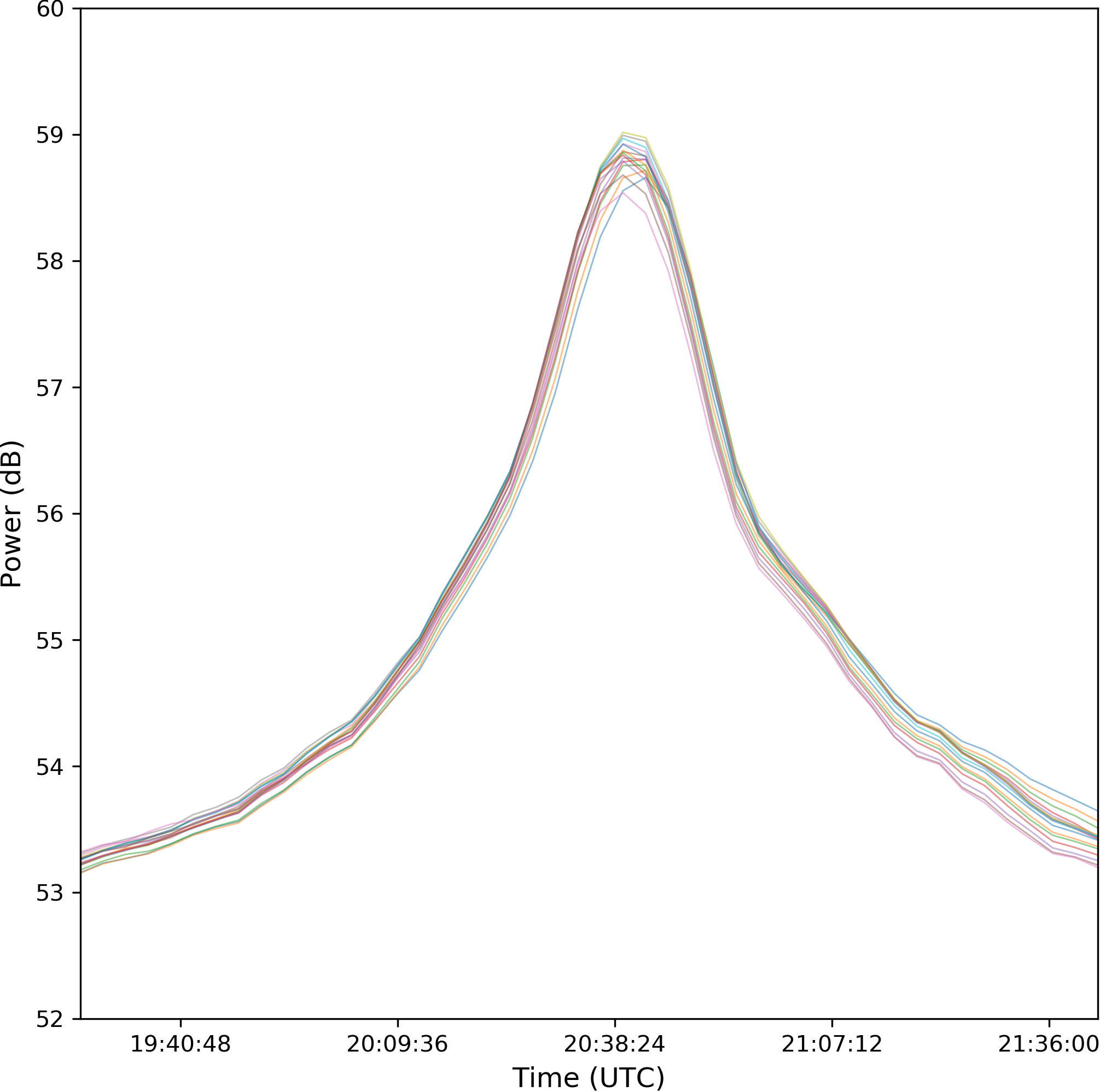}
    \caption{A comparison between the peak retrieved for the galactic core transit after calibrating with an average element pattern (above) and with embedded element patterns (below), using Sagittarius A* itself at different elevations as the calibration source.}
    \label{fig:avg_vs_emb}
\end{figure}

% figure
\begin{figure}
	\includegraphics[width=\columnwidth]{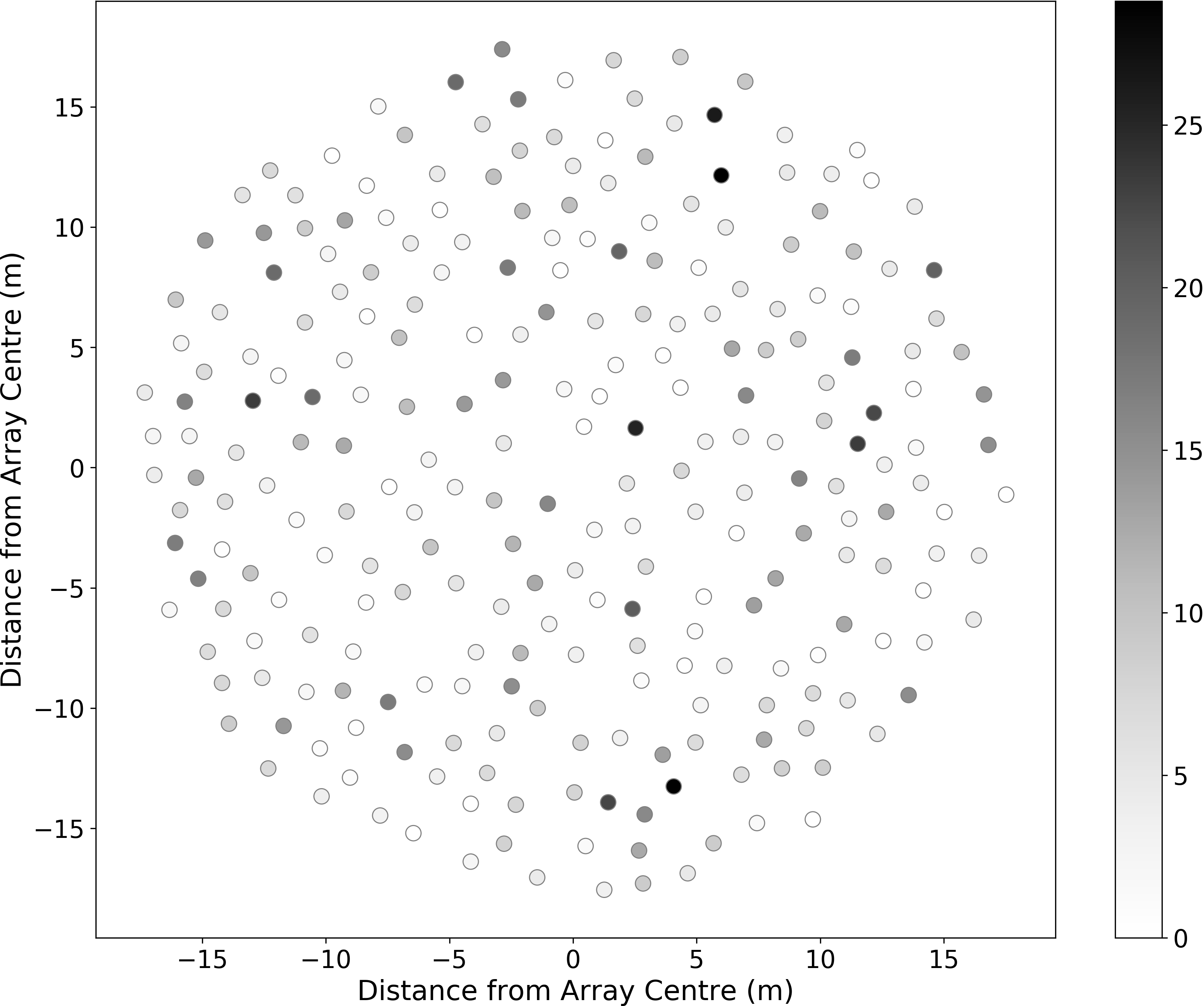}
    \caption{Per-element phase calibration coefficients, in degrees (absolute values), using sky model visibilities generated with an average element pattern to calibrate mock real visibilities generated using embedded element patterns.}
    \label{fig:err_avg_vs_emb}
\end{figure}

\subsubsection{Light Curve Computation}

Validation of observation simulations required computation and assessment of light curves for AAVS-1 over a sufficient observation period. Light curves for a stationary telescope such as AAVS-1 show the total power observed by the array as the Earth rotates over the observation period.

The light curve computation firstly requires the calculation of the station beam \ensuremath{\bar{e}_{S}} as shown in equation \ref{eq:LC_1}, where \ensuremath{e_{H}} and \ensuremath{e_{V}} are the horizontal and vertical embedded response planes, either as calculated in real-time from the spherical wave expansion or loaded from previously computed patterns.

\begin{equation}
    \bar{e}_{S} = \bigg| \sqrt{e_{H}^{2}(\theta, \phi) + e_{V}^{2}(\theta, \phi)} \bigg|
    \label{eq:LC_1}
\end{equation}

The integrated station beam is then calculated by integrating over all sky directions \ensuremath{\theta}, \ensuremath{\phi} as demonstrated in equation \ref{eq:LC_2}.

\begin{equation}
    \bar{e}_{IS} = \int\limits_{0}^{2\pi} \int\limits_{0}^{\pi} \big[ \bar{e}_{S}  \big]^{2} \sin{\theta} d\theta d\phi
    \label{eq:LC_2}
\end{equation}

The directive beam \ensuremath{\bar{e}_{DS}} is computed using the station beam and integrated station beam as shown in equation \ref{eq:LC_3}.
\begin{equation}
    \bar{e}_{DS} = \frac{4\pi \bar{e}_{S}^{2}}{\bar{e}_{IS}}
    \label{eq:LC_3}
\end{equation}

Finally, a single light curve point \ensuremath{L_{\tau}} during an observation at time \ensuremath{\tau} is retrieved following integration over all sky directions \ensuremath{\theta}, \ensuremath{\phi} for the directive beam and the sky \textit{T} visible from the AAVS-1 location at time \ensuremath{\tau}, as demonstrated in equation \ref{eq:LC_4}. This is repeated for all time step intervals of the observation, obtaining light curve values for the entire observation, thus constructing the observation light curve.

\begin{equation}
    L_{\tau} = \int\limits_{0}^{2\pi} \int\limits_{0}^{\pi} \Big[ \bar{e}_{DS} (\theta, \phi) \cdot T(\theta, \phi) \Big] \sin{\theta} d\theta d\phi 
    \label{eq:LC_4}
\end{equation}

It should be noted that the simulator light curve computation can be requested as a standalone simulator operation, if requested as such by the user.

\subsection{Observation with AAVS-1} 
\label{data_acq}
A 24-hour test observation with AAVS-1 was acquired, starting on 11th April 2019 at 07:12 UTC, using the data acquisition software developed for this prototype, described in \cite{daqmagro}. A single frequency channel was selected to be transmitted to the processing server, which was then saved to disk for offline processing. Channel 204 (159.375 MHz) was selected since it lies within a low Radio Frequency Interference (RFI) part of the band, making it ideal for calibration testing in this case study. 

The AAVS-1 digital back-end was instructed to transmit this channel synchronously from all 256 array antennas. These data streams result in a total of $\sim$8 Gbps, which cannot be saved continuously to disk, so a $\sim$1 ms snapshot was persisted every three minutes. Each snapshot was saved as a separate file containing the channelised voltages from all antennas and associated UTC timestamps accordingly. Upon completion of the observation, a cross correlation matrix for each snapshot was generated using an offline correlator. Selected observation cross-correlation matrices, containing the true uncalibrated baseline visibilities, were subsequently used for calibration purposes with corresponding model cross-correlation matrices, separately generated with the visibilities simulator.

\subsection{Calibration}\label{CAL}

StEFCal was implemented as a calibration routine for AAVS-1 in this study. It is a minimization algorithm, aimed at reducing the differences \ensuremath{\delta} between model visibilities and uncalibrated visibilities to an acceptable tolerance level \ensuremath{\delta_{t}} by retrieving best-fit per-element coefficients (\cite{salvini2014stefcal}). The StEFCal algorithm, as implemented for AAVS-1, is summarized in Appendix 1. As such, model baseline visibilities retrieved from the simulator for an observation time \ensuremath{\tau} and uncalibrated baseline visibilities from AAVS-1 acquired at the same observation time were used to retrieve per-element calibration coefficients accordingly using StEFCal. The frequency of such calibration runs will be required according to the stability of the instrumental coefficients retrieved.

\subsubsection{Results}

Assessment of the calibration routine was initially performed using simulator-generated datasets. Sanity checks using identical model and real mock visibilities as routine inputs yielded an identity matrix \textbf{G} of calibration coefficients, as expected. Additional verification of the routine involved the use of visibilities generated with average element patterns as the calibration model, while using their counterparts generated with embedded element patterns as the mock uncalibrated data. Figure \ref{fig:err_avg_vs_emb} provides the phase coefficients (in absolute degrees) retrieved per antenna on calibration with these datasets, which constitute an indication of the expected variation in coefficients retrieved when calibrating with either embedded or average patterns. A median per-antenna phase coefficient variation of 4.7$^{\circ}$ was found to be introduced when calibrating mock real visibilities as described. As such, this is the median phase coefficient error expected to be introduced through the use of an average element pattern for the purposes of generating model visibilities for calibration.

Calibration quality was subsequently assessed using real AAVS-1 data, obtained via a 24 hour observation as described in Section \ref{data_acq}. Calibration on Sagittarius A* was carried out at different elevation pointing directions, at no lower than 60 degrees altitude. Observation correlation matrices from two hours before and after the meridian transit of Sagittarius A* were thus retrieved. Calibration was subsequently carried out with matching model visibilities, generated with the simulator described in this study. Calibration results were separately retrieved with model visibilities generated using embedded or average element patterns accordingly.

Pointing in different sky directions using calibration solutions and an array steering vector therefore allowed for calibrated observations of different radio sources at different times to be obtained. With the 24 hour data set retrieved from AAVS-1, it was possible to point towards different sources at different times, thus allowing observations of different sources accordingly. 

% figure
\begin{figure*}
	\includegraphics[width=\textwidth]{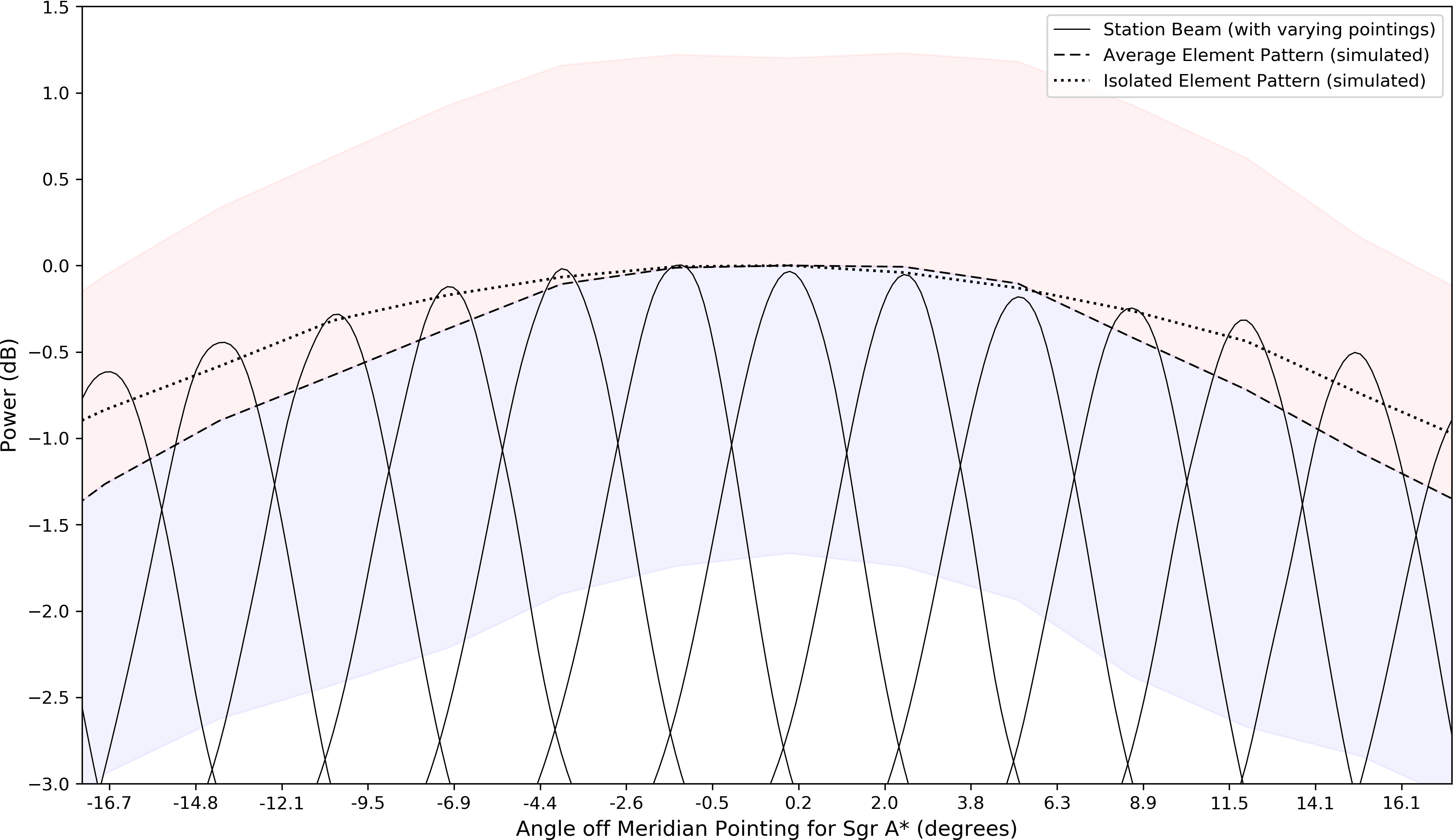}
    \caption{Comparison of normalized peak station beam power retrieved for observations of Sagittarius A* at different elevations, observed over 4 hours with AAVS-1. The simulated average element pattern is included, with a standard deviation range identifying maximum and minimum embedded element pattern values, shown as two highlighted range regions above and below the model average pattern. The figure showcases an increasing residual difference between the observed and simulated average primary beam away from zenith, with the difference being negligible closer to zenith. The isolated element pattern is also included and demonstrates closer agreement to the observed average element pattern.}
    \label{fig:sb_ap_ip}
\end{figure*}

Figure \ref{fig:cen_emb} shows this 24 hour observation, calibrated using model visibilities generated with embedded element patterns pointing towards the meridian transit coordinates for Sagittarius A*. The calibrated observation itself is pointed towards the meridian transit coordinates for Centaurus A. The same sky region towards which the array has been pointed for the observation of Centaurus A is also transited by the Vela-X pulsar and a region of the galactic plane located in Scorpius.

The radio galaxy transits the meridian at 16:20UTC on the observation date. The Vela pulsar, the brightest pulsar at the observation frequency (\cite{manchester2005australia}), transits around 2 degrees off from the same coordinates at 11:27UTC, explaining the observed peak at this time. The largest, final peak observed is the galactic plane itself transiting the same coordinates later on.

% figure
\begin{figure}
	\includegraphics[width=\columnwidth]{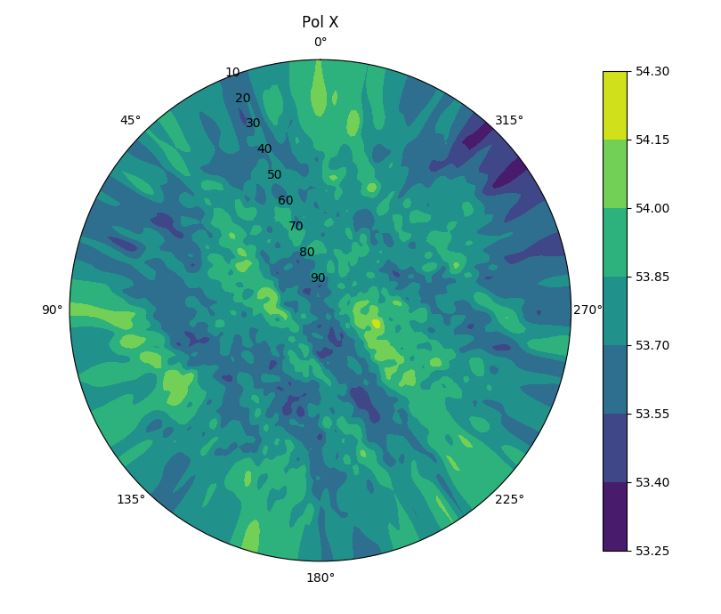}
	\includegraphics[width=\columnwidth]{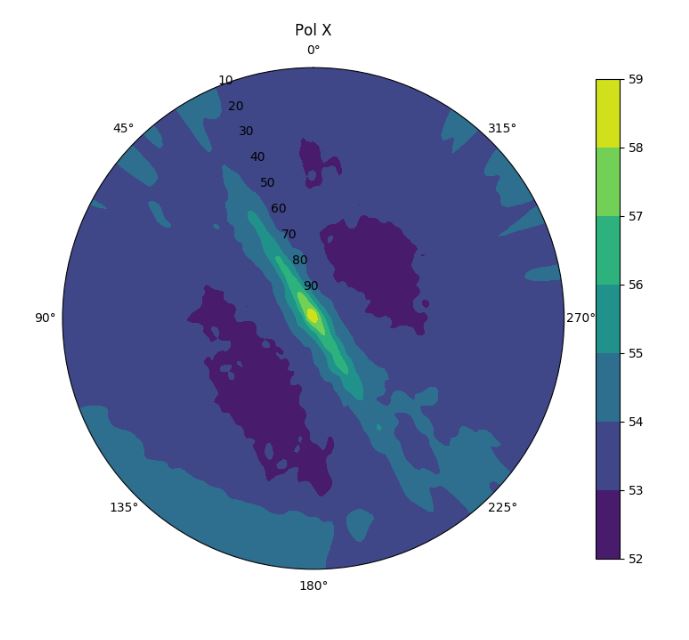}
    \caption{Comparison showing sky power (in dB) retrieved in all sky directions \ensuremath{\theta, \phi} for the X polarization, with data sets before calibration (top) and after calibration (bottom) at the meridian transit time of Sagittarius A*.}
    \label{fig:holog_map}
\end{figure}

Calibration was carried out using uncalibrated and model data sets at matching 15 minute intervals as Sagittarius A* transited the meridian, starting and ending at over 60 degrees elevation. In the first test case, calibration was carried out using model visibilities generated with embedded element pattern responses. It was noted that peak power retrieved for all three transiting sources varied only slightly, with a negligibly stronger peak power retrieved after calibrating with Sagittarius A* at higher elevations, closer to the meridian, as shown in Figure \ref{fig:avg_vs_emb}. 

A similar observation was made when calibrating with model visibilities generated using an average element pattern for the array, also demonstrating negligible variation in calibration quality with different elevations of Sagittarius A*. This is also showcased in Figure \ref{fig:avg_vs_emb}, with calibration coefficients retrieved at varying Sagittarius A* elevation and azimuth positions showing a small residual difference in the Sagittarius A* peak retrieved for simulations with both pattern types. This result seems to indicate that calibration using a model simulated with an average element pattern returns similarly favourable and stable calibration results at different calibration source elevations, when compared with the same calibration routine using a model simulated with embedded element patterns. 

The apparently similar precision in calibration coefficients retrieved in the former case is indeed unexpected, since the models in simulations with embedded patterns are expected to portray a more realistic representation of true mutual coupling effects in the array. This could indicate that true mutual coupling effects are not strong enough to skew calibration results when they are not considered, with consideration of an average element pattern proving sufficient. Such a result can be confirmed in future studies using independent UAV measurements to cross-validate element pattern simulations, therefore allowing for a more complete comparison to be effectively drawn. This result could have a positive impact on calibration routine efficiency and resource requirements, since it would seem unnecessary to consider an embedded element pattern for every antenna in the final SKA1-Low telescope.   

Figure \ref{fig:holog_map} is a holographic representation of total power retrieved by the array in all sky directions \ensuremath{\theta, \phi}, with a pre-calibration data set showing no visible structure as opposed to the post-calibration data set, which shows the galactic plane passing through the meridian. This matches the expected observed sky at the time of retrieval of the imaged data set, shown in Figure \ref{fig:sky_model} as the model sky at the meridian transit time for Sagittarius A*. In the example showcased in the figure, calibration was carried out using visibilities generated with an average element pattern, but an almost identical result was obtained when calibrating with an embedded element pattern.  

Assessment of calibration quality at different elevations was subsequently carried out by calibrating on a strong calibrator (Sagittarius A*) at different times and then retrieving the station beam for every different pointing direction calibrated on, as demonstrated in Figure \ref{fig:sb_ap_ip}. Correct analysis of the varying station beam peak power retrieved at different pointing coordinates requires an understanding of the modulation of the station beam with the primary beam itself. This effect is also required to be well characterized for an accurate comparison of the power received from different sources transiting the meridian at different altitudes. As shown in Figure \ref{fig:sb_ap_ip}, a change in peak station beam power is noted when observing the same source at different elevations - in this case, the galactic core. 

When pointing the station beam towards a radio source as it transits across the sky, the power received from the source is therefore expected to vary with the source altitude, depending on the average primary beam response for the array elements. This expectation labours under the assumption that the average element pattern approaches the isolated element pattern, provided that enough embedded element patterns are averaged. The isolated element pattern is defined as the antenna pattern measured in an isolated infinite plane, ergo experiencing no mutual coupling effects from neighbouring antennas. 

This assumption was verified using selections of progressively larger concentric sub-arrays from the 256 AAVS-1 antennas, starting off with a sub-selection of the central 8 antennas and progressively doubling the size of the array until reaching the full 256 antenna configuration. For each sub-array selection, the corresponding average element pattern was calculated at the horizontal coordinates for the transit of Sagittarius A*. This was done in order to be able to use a real observation of the galactic core's passage to verify and compare average pattern computation directly. As expected, it was observed that the model average element pattern approached the model isolated element pattern with an increasing array size. The simulated average pattern with 256 antennas is included in figure \ref{fig:sb_ap_ip}.

Figure \ref{fig:sb_ap_ip} demonstrates the station beams retrieved for observations of Sagittarius A* carried out at different elevations over 4 hours. The station beams are displayed with a -3.0 dB cut-off, equivalent to a decrease by half in power from the highest station beam peak, normalized to 0dB. The model isolated element pattern and the model average primary beam are also displayed on the same figure. The latter was calculated by taking all 256 embedded element patterns into account, sampling power at the same discreet horizontal coordinates at which Sagittarius A* was observed. 

It was noted that the average element pattern retrieved from the observation of the passage of the galactic core at different elevations was closer to the isolated element pattern than the model average element pattern. The retrieved average pattern can be determined by measuring peak station beam power at different elevations, which can be calculated from the station beams computed in Figure \ref{fig:sb_ap_ip}. This calculated average element pattern, using the observation of Sagittarius A*, was found to be slightly wider than both the isolated element pattern and the model average element pattern, meaning that larger deviations between the observed and model average patterns were observed farther away from the quasi-zenithal pointing for the meridian transit of the galactic core. 

Indeed, measurements recorded <5 degrees off the meridian pointing elevation for Sagittarius A* exhibit negligible differences between the retrieved average pattern and the modeled average or isolated primary beams. This is in contrast with measurements taken $\sim$15 degrees off the same meridian pointing elevation, with a difference of $\sim$0.2dB between the retrieved average pattern and the isolated element pattern. The difference recorded increases to $\sim$0.7dB when comparing the retrieved average pattern against the model average element pattern. These differences could be attributed to minor inaccuracies in calibration solutions or geometric pointing errors. It should also be noted that power inadvertently received from prevalent sidelobes could also result in an increased power retrieved for different array pointing coordinates. This could thus explain why the retrieved average pattern from different station beam pointing coordinates is slightly broader than both the modeled average element pattern and the isolated element pattern, with pointings further away from zenith showing an increased divergence from expected power.

\section{Conclusions and Further Work}

This study explored the possibility of using embedded pattern responses to simulate baseline visibilities for calibration of an AAVS-1 station for complete consideration of mutual coupling effects in such a phased array. A simulator capable of producing such model visibilities for calibration was implemented, capable of using either embedded pattern responses or average pattern responses as required accordingly. Model visibilities generated using the simulator were then used for calibration with StEFCal. The primary objective of calibrating AAVS-1 with the implemented simulator and calibration routine was successfully carried out, providing one possible strategy for calibrating a SKA1-Low station. Observations of Sagittarius A* and Centaurus A, amongst others, were successfully carried out using a 24 hour observation with AAVS-1.  

No significant difference between the use of embedded or average element pattern responses was noted when calibrating using StEFCal, with neither demonstrating better coefficient precision with changing calibration source elevation. It was thus observed that on-sky calibration with sky model visibilities generated with both pattern types was similarly successful. This observation can be of particular importance for future calibration observations and routines for AAVS-1 and possibly similar SKA1-low stations, particularly regarding the assessment of the necessity for distinct element patterns to be stored and used for model visibilities simulation across SKA1-Low.

Indeed, with a negligible difference observed in calibration results retrieved with either pattern type, it can be argued that the use of an average element pattern for an AAVS-1 type station of 256 antennas would reduce operational and memory costs for calibration. It must be noted that retrieving embedded patterns using the spherical harmonic wave expansion results in a computational cost penalty, while retrieving pre-computed embedded patterns would necessitate a memory penalty. The efficiency of calibrating with an average element pattern would thus result in faster coefficient retrieval, of importance for per-channel real-time calibration of such a SKA1-Low station. It must be noted that while this observation holds true for AAVS-1, observations using other similar stations would need to be carried out to verify that such behaviour is consistent across stations. The observations made in this regard in this study can be confirmed in future work with independent in-situ measurements of the embedded element patterns to validate their simulated counterparts used in this study.

The calibration routine architecture developed here also allows for future implementations of calibration methods other than StEFCal to be tested out and compared against calibration results with StEFCal accordingly. This would eventually allow a selection of calibration methods which can be user-defined for a particular calibration observation accordingly. With the appropriate model visibilities generated using the implemented simulator, such a calibration routine would constitute a good prototype for calibrating SKA1-Low frequency aperture array stations, on a per-frequency-channel basis, in real time. 

\section{Appendices}

\subsection{Appendix 1: StEFCal}

StEFCal can be showcased as shown hereunder, reproduced from \cite{salvini2014stefcal}.
\\

\begin{algorithmic}
\STATE \textbf{Initialization: }\ensuremath{G^{[0]} = \textbf{I}}
\FOR{i in 1 .. maxiter}
    \STATE $G^{i} = G^{i-1}$
    \FOR{p in 0 .. nelem}
        \STATE $Z_{:,p} = G^{i} \cdot \big \langle M_{:,p} \big \rangle$
        \IF{${Z_{:,p}^{H} \cdot Z_{:,p}} \neq 0$}
            \STATE $g_{p} = \big \langle V_{:,p}\big \rangle \cdot Z_{:,p}/Z_{:,p}^{H} \cdot Z_{:,p}$
        \ENDIF
        \STATE $G^{i} = g$
        \IF{$||G^{i} - G^{i-1}||_{F}/||G^{i}||_{F} \leq \delta_{t}$}
            \STATE \textbf{break}
        \ENDIF
    \ENDFOR
\ENDFOR
\end{algorithmic}

The complex coefficient solutions \textbf{G} are updated after every iteration \textit{i}, and the algorithm will proceed to the next iteration provided that the ratio of the Frobenius norm of the difference between the coefficients retrieved in the current and previous iterations and the Frobenius norm of the coefficients retrieved in the current iteration is not less than a tolerance level \ensuremath{\delta_{t}}. If this ratio falls below the user-defined \ensuremath{\delta{t}}, convergence is reached. Calibration is terminated either on convergence or if the user-defined number of maximum algorithm iterations, \textit{maxiter}, is reached before convergence. In the latter case, the user is prompted that convergence has failed accordingly.

The complex coefficients are initialized as an identity vector \textbf{I}, updated after every iteration respectively. The latest coefficients retrieved from a previous iteration are used to perturb the model visibilities \ensuremath{\big \langle M_{:,p}\big \rangle}, in order to approach the observed visibilities \ensuremath{\big \langle V_{:,p}\big \rangle}. New estimates of \textbf{G} are calculated for every element \textit{p} by considering all baselines involving element \textit{p} and solving \textit{nelem} linear least squares problems per iteration (per element) accordingly.

In terms of the model sky visibilities, \textit{M}, and the observed sky visibilities, \textit{V}, the measurement equation can be defined as shown in equation \ref{eq:STE_1}. The diagonal gain matrix, \textit{G}, refers to the per-element errors which perturb the model to reproduce the actual observed visibilities.

\begin{equation}
    V = G^{H} M G
    \label{eq:STE_1}
\end{equation}

In order to estimate the diagonal values of the gain matrix \textit{G}, it will be necessary to minimize the Frobenius norm of the difference between the perturbed model and the observed visibilities, as demonstrated in equation \ref{eq:STE_2}.

\begin{equation}
    \min \Big|\Big| V - G^{H} M G \Big|\Big|^{2}_{F}
    \label{eq:STE_2}
\end{equation}

In order to initialize the minimization, the first gain matrix \ensuremath{G^{H}} is taken as the identity matrix. The Frobenius norm is the sum of the squared values for the baseline visibilities pertaining to every antenna \textit{i}. It is thus possible to consequently sum over the Frobenius norms for all elements, and therefore the minimization problem can be written as in equation \ref{eq:STE_3}, where \ensuremath{|| e ||_{F}} is the Frobenius norm of the error after a minimization iteration. 

\begin{equation}
    \sum\limits_{p}^{N} \Big|\Big| V_{:,p} - (I^{H} M G_{:,p}) \Big|\Big|^{2}_{F} = \Big|\Big| e \Big|\Big|_{F}
    \label{eq:STE_3}
\end{equation}

The terms \ensuremath{I^{H}} and \textit{M} are both known, and can be collectively identified as \textit{Z}. Additionally, since \textit{G} is a diagonal matrix, it is also possible to redefine \textit{G} as a vector \textit{g} containing the diagonal values only. In this manner, equation \ref{eq:STE_3} can be rewritten as in equation \ref{eq:STE_4}.

\begin{equation}
    \sum\limits_{p}^{N} \Big|\Big| V_{:,p} - (Z_{:,p} \cdot g_{p}) \Big|\Big|^{2}_{F} = \Big|\Big| e \Big|\Big|_{F}
    \label{eq:STE_4}
\end{equation}

Consequently, the minimization can be carried out by minimizing the sum of the Frobenius norms for the visibilities for every antenna's visibilities. This will allow for the calculation of every \ensuremath{g_{p}} term for every antenna \textit{p} as shown in equation \ref{eq:STE_5}, using the normal equations method. 

\begin{equation}
    g_{p} = \big[Z^{H}_{:,p} \cdot Z_{:,p}\big]^{-1} V_{:,p} \cdot Z^{H}_{:,p}
    \label{eq:STE_5}
\end{equation}

Equation \ref{eq:STE_5} can be re-written as shown in the algorithm minimization computation, demonstrated in equation \ref{eq:STE_6}.

\begin{equation}
    g_{p} = \frac{V_{:,p} \cdot Z^{H}_{:,p}}{Z^{H}_{:,p} \cdot Z_{:,p}} 
    \label{eq:STE_6}
\end{equation}

In this manner, the new estimates \textit{g} for the error contributions per antenna are retrieved. In the next minimization iteration, these coefficient estimates will replace the identity matrix of initialization coefficients \ensuremath{I^{H}}, and the minimization proceeds accordingly for every iteration. The Frobenius norm of the change in the gain  estimates \textit{g} between the current iteration and the previous iteration is used as a convergence criterion. When the change is below a user defined \ensuremath{\delta_{t}}, it is assumed that the retrieved \textit{g} sufficiently minimizes the difference between the observed and model visibilities, and therefore accounts for the true per element errors. This convergence criterion estimation is shown in equation \ref{eq:STE_7}.

\begin{equation}
    \delta_{t} = \frac{\Big|\Big|G^{i} - G^{i-1}\Big|\Big|_{F}}{\Big|\Big|G^{i}\Big|\Big|_{F}} 
    \label{eq:STE_7}
\end{equation}

%%%%%%%%%%%%%%%%%%%%%%%%%%%%%%%%%%%%%%%%%%%%%%%%%%



\bibliographystyle{mnras}
\bibliography{paper} % if your bibtex file is called example.bib

%\begin{thebibliography}{99}

%%%%%%%%%%%%%%%%%%%%%%%%%%%%%%%%%%%%%%%%%%%%%%%%%%

%%%%%%%%%%%%%%%%% APPENDICES %%%%%%%%%%%%%%%%%%%%%

%%%%%%%%%%%%%%%%%%%%%%%%%%%%%%%%%%%%%%%%%%%%%%%%%%

% Don't change these lines
\bsp	% typesetting comment
\label{lastpage}
\end{document}